%% file: AxionFlavor_3.tex
\documentclass[aps,prd,
twocolumn,superscriptaddress,nofootinbib,10pt,preprintnumbers,longbibliography
]{revtex4-1}

\usepackage{blkarray}

\usepackage[colorlinks=true,linkcolor=black,citecolor=blue,urlcolor=blue, pdfborder={0 0 0}]{hyperref}
\usepackage{mathtools}
\usepackage[utf8]{inputenc}
\usepackage{color}
\usepackage[table,xcdraw,dvipsnames]{xcolor}
\usepackage{multirow}
\usepackage[normalem]{ulem}    
\usepackage{mathtools}
  \usepackage{amsmath}
  \usepackage{amssymb}
  
    \usepackage{calrsfs}    
  \DeclareMathAlphabet{\mathcal}{OMS}{zplm}{m}{n}   
  
\definecolor{ultramarine}{RGB}{0,32,96}
\definecolor{blue-violet}{rgb}{0.7, 0.2, 0.8}

\newcommand{\mX}{\mathcal{X}}

\newcommand{\mA}{\mathcal{A}} 
 
\newcommand{\mU}{\mathcal{U}} 
\newcommand{\mV}{\mathcal{V}} 
\newcommand{\mY}{\mathcal{Y}}
\newcommand{\mM}{\mathcal{M}}

\newcommand{\UPQ}{U(1)_{\mathrm{PQ}}}

      \def\vev#1{\langle #1\rangle}

\def\Tr{\mbox{Tr}\,}

\def\det{\mbox{det}\,}

\newcommand{\GF}{{\mathcal{G}_F}}

\newcommand{\beq}{\begin{equation}}
\newcommand{\eeq}{\end{equation}}
\newcommand{\bea}{\begin{eqnarray}}
\newcommand{\eea}{\end{eqnarray}}

\newcommand{\eqn}[1]{Eq.~(\ref{#1})}

\def \gsim{\mathrel{\vcenter
     {\hbox{$>$}\nointerlineskip\hbox{$\sim$}}}}
     
\def\subPQv{{\begin{picture}(12,0)(0,0)\put(0,0){$\scriptscriptstyle \rm
PQ$}\put(-.9,.5){\line(4,1){10}}\end{picture}}}

\begin{document}

\title{The axion flavour connection}

\newcommand{\affCNRS}{{\small \it
Institut de Physique des 2 Infinis de Lyon (IP2I), UMR5822, CNRS/IN2P3, F-69622 Villeurbanne Cedex, France}} 

\newcommand{\affINFN}{{\small \it INFN, Laboratori Nazionali di Frascati, C.P.~13, 100044 Frascati, Italy}}

\newcommand{\affSISSA}{{\small \it  SISSA International School for Advanced Studies, Via Bonomea 265, 34136, Trieste, Italy}} 
\newcommand{\affINFNTS}{{\small\it INFN - Sezione di Trieste, Via Bonomea  265, 34136, Trieste, Italy}}

\author{Luc Darm\'e}
\email{l.darme@ip2i.in2p3.fr}
 \affiliation{\affCNRS}
\author{Enrico Nardi}
\email{Enrico.Nardi@lnf.infn.it}
\affiliation{\affINFN}
\author{Clemente Smarra}
\email{csmarra@sissa.it}
\affiliation{\affSISSA}\affiliation{\affINFNTS}

\begin{abstract}
A  local flavour symmetry acting on the quarks of the Standard Model  
can automatically give rise  to an accidental global  $U(1)$   which remains preserved 
from sources of explicit breaking up to a large operator dimension,
while it  gets spontaneously broken together with the flavour symmetry.  
Such non-fundamental symmetries  are often endowed  with a mixed QCD anomaly, 
so that  the strong CP problem  is automatically solved via the axion 
mechanism. 
We  illustrate the general features required to realise this scenario, 
and  we discuss a simple construction based on the flavour group 
$SU(3)\times SU(2) \times U(1)_F$  to illustrate how mass hierarchies can arise 
while ensuring at the same time a high quality Peccei-Quinn symmetry. 
\end{abstract}

  \maketitle

 \parskip 2pt

\section{Introduction} 
\label{sec:intro}
Fermion family replication suggests that quarks belong to representations 
of some non-Abelian symmetry group, while the absence of degenerate multiplets
in the mass spectrum  hints to  spontaneous  breaking (SB) of the symmetry. Identifying 
the structure of a hidden symmetry  just from  data grounded in the broken phase 
can be an awfully difficult  task. Complying with some theoretically well motivated    
principles might thus provide a  crucial guide to advance in this  endeavour. 
Here we  put forth the idea that a  flavour\footnote{In this 
work `flavour' refers  to a replication of the known quarks as well as of new exotic quarks.} gauge symmetry  $\GF$  acting on particles carrying colour, 
as well as on certain scalar multiplets responsible for  the SB of $\GF$, automatically enforces the invariance 
under a global $U(1)$ of  all gauge invariant  operators constructed with  quarks and 
scalar fields, up to some suitably large dimension.  Such a symmetry,  being non-fundamental,  
can be  endowed with a mixed $U(1)$-$SU(3)_{\rm QCD}$ anomaly,  and being broken  
spontaneously by the same vacuum expectation values (VEVs) that break $\GF$,   
will  give rise to an axion~\cite{Peccei:1977hh,Peccei:1977ur,Weinberg:1977ma,Wilczek:1977pj}. 
Local flavour groups of this type, if they exist, will thus automatically provide a solution to 
the strong CP problem~\cite{Callan:1976je,Jackiw:1976pf} via the Peccei-Quinn (PQ) mechanism.

It is well known that in benchmark axion models, like the    
Kim-Shifman-Vainshtein-Zakharov (KSVZ)~\cite{Kim:1979if,Shifman:1979if}  and  Dine-Fischler-Srednicki-Zhitnitsky 
(DFSZ)~\cite{Zhitnitsky:1980tq,Dine:1981rt}   models, in which the axion is mainly hosted in a 
complex scalar gauge singlet $\Phi$,  the origin of the PQ symmetry remains misterious. This is because  
there is no  reason to forbid renormalizable operators like  $\mu^3 \Phi, \mu^2\Phi^2,\dots$  
since they do not violate any fundamental principle.   However,  if present, such operators would destroy the PQ 
symmetry together with the axion solution.   
A satisfactory explanation  of the {\it origin} of the PQ symmetry would arise if,  
given a suitable extension of the Standard Model (SM),   all renormalizable Lagrangian terms respecting 
first principles (Lorentz and local gauge invariance) automatically  preserve also a global $U(1)$ with 
the required properties. Moreover, in order to solve the strong CP problem, $\UPQ$ must be of a very high {\it quality}, 
in the sense that any source of  explicit breaking  (besides the QCD anomaly) must be extremely suppressed. 
This results in the requirement that  PQ breaking effective operators of dimension $D\lesssim 10$
should remain forbidden~\cite{Dine:1986bg,Barr:1992qq,Kamionkowski:1992mf,Holman:1992us,Ghigna:1992iv}. 
Different approaches have been put forth to tackle the PQ origin and quality problems.
Composite models assume that there are no axion-related fundamental scalars. In this case 
besides local symmetries the PQ quality is also assisted by Lorentz invariance, which helps to raise the dimension 
of  PQ symmetry breaking operators~\cite{Randall:1992ut,Redi:2016esr,Lillard:2018fdt,Lee:2018yak,Gavela:2018paw,Vecchi:2021shj,Contino:2021ayn}. 
If instead the axion sits in one or more fundamental scalars, then 
to generate and protect $\UPQ$ up to some operator dimension $D$, one can  rely only on local symmetries. 
Discrete gauge symmetries $\mathbb{Z}_D$~\cite{Krauss:1988zc,Chun:1992bn,Dias:2002gg,Carpenter:2009zs,
Harigaya:2013vja,Dias:2014osa,Harigaya:2015soa,Ringwald:2015dsf},
Abelian gauge symmetries with multiple complex  scalars with values of  the gauge charges of order $D$~\cite{Barr:1992qq}, 
non-Abelian local symmetries generally  of degree not less than~$D$~\cite{DiLuzio:2017tjx,Ardu:2020qmo,Yin:2020dfn},  
gauge symmetries  assisted by  supersymmetry~\cite{Lillard:2017cwx,Nakai:2021nyf,Choi:2022fha}  or by higher dimensional constructions~\cite{Hill:2002me,Hill:2002kq,Yamada:2021uze} have been used for this scope.\footnote{Recently, it has also 
been argued that the axion quality problem can be  automatically 
solved  in an   alternative  formulation in which the axion is introduced as a 2-form $B_{\mu\nu}$,  
whose Lagrangian enjoys  an invariance under a gauge shift   $B_{\mu\nu}\to  B_{\mu\nu} +  \Omega_{\mu\nu}$
 which protects the axion mechanism~\cite{Dvali:2022fdv}.}
There are some unsatisfactory aspects  common to most of these solutions. For example they generally introduce  
a completely new sector, which is functional to generate $\UPQ$ but is otherwise loosely connected to other SM properties.
Moreover, as regards the level of protection, it is often controlled by some ad hoc feature which 
involves numbers of  order $D$   (the degree of a gauge group, the value of Abelian charges, etc.)

In Ref.~\cite{Darme:2021cxx}  it was pointed out that a certain type of 
semi-simple local symmetries
acting on a suitable set of scalar multiplets and of fermions carrying colour, are well suited to generate and protect 
a $\UPQ$,  even when  each group factor in the semi-simple decomposition 
has degree much smaller than $D$. 
This suggests the interesting possibility of enforcing  
accidentally a PQ  symmetry by exploiting  group factors  of degree  $\leq 3$, which can thus be interpreted
as  flavour symmetries acting on the SM quark generations. 
 In this work, we study the viability of such a scenario, we state some necessary conditions for its realisation  
and, as an example,  we  construct  a flavor model 
based on a particularly simple flavour group in which a global $\UPQ$  arises accidentally and 
remains protected up to operator dimension $D=10$, while mass hierarchies compatible with 
those observed in the quark sector are generated upon minimization of a scalar potential.
The paper is organised as follows.
In Section~II we  state some conditions which must be matched by 
theoretical constructions that aim to realise this idea. In Section~III we describe 
the main steps through which flavour models  of this type can be constructed. In 
Section~IV we present a concrete realisation, and finally  in Section~V we summarise the main 
results and draw our conclusions. 

\section{The need for an Abelian  factor} 
\label{sec:abelian}

An important step towards an explanation of the origin of the axion, is the search for 
a scalar potential that respects automatically a global $U(1)$. 
A  class of semi-simple  symmetries that, when imposed as local symmetries
 acting on a certain set of scalar multiplets,  features  particularly interesting 
properties  to achieve this goal, was studied in  Ref.~\cite{Darme:2021cxx}. 
They have the generic structure $\GF = SU(m)\times SU(n)$ with $m\neq n$. 
For example the potential for a scalar multiplet 
transforming in the bi-fundamental $Y\sim (m,\bar n)$  features an {\it exact} accidental $U(1)$ 
corresponding to a global rephasing $Y\to e^{i\alpha} Y$. This  follows from the fact that 
all   gauge invariant operators, e.g. ${\rm Tr}(Y^\dagger Y),\; {\rm Tr}(Y^\dagger Y Y^\dagger Y)$,
are Hermitian,   since for $m\neq n$ the would-be non-Hermitian determinant operator $\sim \epsilon^{\alpha_1 \alpha_2 \dots }
\epsilon_{i_1i_2\dots } Y_{\alpha_1}^{i_1}  Y_{\alpha_2}^{i_2}  \dots $ vanishes identically. 
This is a promising observation; however,  a model aiming to provide a satisfactory explanation of   
the origin of the PQ symmetry needs to satisfy additional requirements, the most important of which are:
\begin{itemize} \itemsep -1pt
\item[(i)] The accidental $U(1)$ symmetry enforced by $\GF$ must have a colour anomaly;  
\item[(ii)] All  fermions carrying colour   must be massive;
\item[(iii)]   $U(1)$ must remain preserved up to operators of dimension 
$D\gsim 10$ without introducing  a too large 
number of chiral  quarks.
\end{itemize}
While the potential  of a  single scalar multiplet transforming under 
a {\it rectangular} gauge group does satisfy the quality requirement~\cite{Darme:2021cxx},  
its Yukawa sector unavoidably yields some massless quarks and fails to  
satisfy the requirement (ii). Thus, additional scalars must be introduced. 
In general,  this opens up the possibility of writing also non-Hermitian operators, 
and then the  issue whether a global $U(1)$ does arise, and up to which 
operator dimension it remains preserved, becomes non-trivial.  
The main reason for the last  requirement (iii) is to  improve with respect to existing models.  
 For example,  the   model discussed in Ref.~\cite{DiLuzio:2017tjx},  which is  based on the 
gauge symmetry  $SU(D)_L\times SU(D)_R$, contains  exotic quark multiplets trasforming 
as $\mathcal{Q}_L  \sim (D,1)$  $\mathcal{Q}_R  \sim (1, D)$  coupled to a scalar multiplet  
$Y\sim (D,\bar D)$.  This model  satisfies (i) and (ii)  while the first PQ-breaking operator 
($\det Y$) arises at dimension $D$. Here we seek more economical constructions  
involving fermion representations of dimension much smaller than $D$, 
 suitable to host the SM quarks.

We will now prove that semi-simple gauge groups $\GF$ 
do not suffice  to enforce an accidental $U(1)$ symmetry while 
satisfying simultaneously  the three requirements (i)-(iii).
Let us consider a generic semi-simple gauge group $\GF= \left[\Pi_\ell SU(M_\ell)\right]_L \times \left[\Pi_r SU(N_r)\right]_R$ 
 acting on a certain set of scalar multiplets $Y^{\ell r}\sim (m_\ell, \bar n_r)$  in the fundamental ($m_\ell=M_\ell,\; n_r = N_r$) or trivial 
($m_\ell,\,n_r=1$) representation of pairs of  group factors, and 
a set of left-handed (LH) and right-handed (RH) quarks $Q_L\,, Q_R$ of a given electric charge 
in the fundamental (or trivial) representation of a single group factor.\footnote{We assume
for definiteness  that all fields transform in the fundamental or trivial representation of the group factors. Because of superselection rules 
considering one quark sector of a given electric charge
is without loss of generality.}
To prevent a QCD gauge anomaly and to ensure that each LH quarks has a  RH chiral partner, so 
that no quark is prevented from acquiring a mass, we require $n_Q \equiv \sum_\ell m_\ell=\sum_r n_r $.
Let us consider  the  set of gauge invariant  Yukawa terms:
\beq
\label{eq:yuk}
\sum_{\{\ell r\}} \eta^{\ell r} \bar Q_{L \ell} Y^{\ell r} Q_{R r} \,, 
\eeq
where $ \eta^{\ell r} $ are coupling constants.\footnote{Here  
$\{\ell r\}$
are not matrix indices. They label the specific 
coupling that multiplies  the Yukawa operator, as well as the particular 
fermion and scalar multiplets involved.  In case  $Q_{L \ell}, Q_{R r} $ belong to 
the same representation under $\GF$  and  under  the $SU(2)_L$  electroweak group, 
then  $(\eta\, Y)^{\ell r} $   represents an invariant mass term.
If instead one $Q_{L,R}$ is a $SU(2)_L$ doublet and the other one a singlet,  then $Y^{\ell r} $  represents a Higgs doublet.} 
Upon SB of $\GF$    ($\Tr Y^\dagger Y  \to  \vev{ \Tr Y^\dagger Y} \neq 0 $)
the Yukawa operators 
generate a mass matrix for the quarks. 
By arranging  the LH  and RH fermions 
in two vectors $\Psi_{L,R} $,   the mass term  can be written as:
\beq
\label{eq:mass} 
\bar \Psi_L \; \mathcal{M}(\eta\,\vev{Y})\;  \Psi_R\,,
\eeq
with $\mathcal{M}$ a $n_Q\times n_Q$  square matrix. Although this matrix can have one or 
more vanishing $m_\ell\times n_r$  blocks (corresponding for example to the absence of 
a  $Y\sim (m_\ell, n_r)$ scalar multiplet) the requirement that all the quarks are massive implies  
 $\det \mathcal{M}\neq 0$. Then the sum 
\beq
\label{eq:dimn}
\epsilon_{i_1 i_2\dots i_{n_Q}}  \epsilon_{j_1 j_2\dots j_{n_Q}} 
\mathcal{M}_{i_1 j_1}
\mathcal{M}_{i_2 j_2} \dots 
\mathcal{M}_{i_{n_Q} j_{n_Q}} 
\eeq 
must contain at least one non-vanishing term. 
Suppressing the coupling constants ($\eta\to 1$)  
so that  each block $\mathcal{M}_{\ell r}$ is replaced by $ \vev{Y}_{\ell r}$),
\eqn{eq:dimn}  ensures  that one can write down  a gauge invariant scalar operator  of dimension $n_Q$ 
whose VEV does not vanish.  Let us now denote by $\mX_\psi$ the 
charge of the field $\psi$ under the accidental $U(1)$.  
From \eqn{eq:mass} we see  that this operator 
must carry a global  charge equal to $\sum_{Q_{L\ell}}  m_\ell \,
\mX_{Q_{L\ell}} - \sum_{Q_{R r}}  n_r \,
\mX_{Q_{R r}} =\mathcal{A}$,  
where $\mathcal{A}$ is the   $U(1)$-QCD  anomaly coefficient.\footnote{Strictly speaking 
$\mathcal{A}$ is the contribution to the anomaly coefficient of one sector of 
quarks with the same electric charge. Applying this argument  sector by sector one reaches 
the same conclusion.}
 We are thus left with the following options: (i)  
$\mathcal{A} =0$: the operator does not break the accidental symmetry, however,  
the symmetry  is non-anomalous and hence it is not  a PQ symmetry; (ii)  $\mathcal{A} \neq 0$:   
$U(1)$ can be promoted to a PQ symmetry  which, however, is unavoidably  
 broken at dimension $n_Q$ by the operator in \eqn{eq:dimn}.  
 Clearly this  brings back  the issue that to enforce the required    $U(1)_{PQ}$ protection
one needs   to  introduce  $n_Q \geq  D$  chiral quarks for any given  electric charge. 
 
 This unpleasant  result can be  circumvented by extending the flavour symmetry 
 to include an Abelian gauge factor  $\GF \to \GF \times U(1)_F$   such that 
 the Yukawa operators in \eqn{eq:yuk} are $U(1)_F$-invariant   and thus  allowed, 
 while the scalar operator corresponding to \eqn{eq:dimn}  is not   $U(1)_F$ invariant 
 and remains forbidden.  The search for suitable $U(1)_F$ symmetries that can 
 realise this picture  plays a central role in the present study.

 
 \section{Model Building}
 \label{sec:modelbuilding}
 
 In this section we describe a general  strategy to   
  construct models  that can realise an  axion-flavour connection.
With the exception of the top mass,  all other SM quark masses 
have values well below the  scale of the  electroweak (EW) VEV $v\sim 246\,$GeV, that is 
the dimensional scale to which all SM masses should be related. 
 This observation suggests the possibility that 
the SM fermion masses could be forbidden at the renormalizable level because
of  some flavour symmetry,  and that they could arise from effective operators that are switched on 
after   the flavour symmetry gets spontaneously broken.  This scenario requires the introduction 
of a certain set of exotic quarks with electric charge $+2/3$ and $-1/3$, 
and of SM singlet scalars transforming in some representation of the flavour group 
to generate the effective operators, 
and to break spontaneously the flavour symmetry. 
In each charge sector the determinant of the $n_Q\times n_Q$ mass matrix 
should  have a parametric dependence   $\det(\mathcal{M})\sim v^3 \Lambda^{n_Q-3}$,
where $\Lambda$ represents some large mass scale unrelated to the EW VEV $v$.  
Note that requiring that there are no massless 
quarks ($\det(\mathcal{M})\neq 0$) already imposes some relevant constraints 
on the viability of the model.  Let us clarify these points with an example.  

 \subsection{A simple example}
 \label{sec:example0}

Let us consider for definiteness the up-quark sector, 
leaving understood that the same construction is replicated in the down-quark sector.   
The simplest flavour symmetry of the form  $\GF \times U(1)_F$ where $\GF$ denotes a semi-simple 
 {\it rectangular}  group,  corresponds to  $\GF = SU(3)\times SU(2)$.
Let us assume that the LH quark doublets $q_L$ transform in the fundamental of $SU(3)$ 
while the RH quarks 
$u_R$ belong to a fundamental  of $SU(2)$. Then one RH quark $t_R$ is a singlet under $\GF$.
To render $\GF$ anomaly free, the simplest choice is to `mirror' this quark content by introducing 
a minimal set of  exotic  quarks $(U_R, U_L, T_L)$. 
Under  $ SU(3)\times SU(2)$ of flavour the quarks transform as
\begin{align}
\nonumber
    &q_L\sim (\bold{3},\bold{1}), \quad u_R\sim (\bold{1},\bold{2}), \quad t_R \sim (\bold{1},\bold{1}) \,, \\
    &U_R \sim (\bold{3},\bold{1}), \quad U_L \sim (\bold{1},\bold{2}), \quad T_L \sim (\bold{1},\bold{1})  \,.
\label{eq:fermions}
\end{align}
Note that while $q_L=(u,d)_L^T$  is an EW doublet,  
the other five up-type quark multiplets are $SU(2)_L$  singlets.
Five  down-type  $SU(2)_L$ singlets $d_R,b_R,D_R,D_L,B_L$ complete the  
fermion content of the model,  but they are irrelevant for the following discussion.
Let us now introduce (besides the Higgs)  the following scalar multiplets:
\begin{align}
\label{eq:scalars}
    & Y\sim (\bold{3},\bar{\bold{2}}), \quad Z\sim (\bold{3},\bold{1}), \quad X \sim (\bold{1},\bar{\bold{2}}).
\end{align}
Although the multiplet $X$, together with the Higgs, would suffice to 
provide masses for all the quarks,  to generate hierarchical  masses 
other fields transforming like  $Y$ and $Z$  are  also 
needed, so we introduce them from the start. 
Denoting with lower case letters the VEVs of the components of the 
scalar fields, e.g. $\vev{X} = (x_1,x_2)$  etc., 
and choosing the field basis in which $\vev{Y_1^1}=y_1,\, 
 \vev{Y_2^2}=y_2$ with all other entries vanishing, 
the  $6\times 6$ mass matrix consistent with invariance under $\GF$
reads:
\begin{align}
\label{eq:66Matrix}
\mathcal{M}=\quad 
\begin{blockarray}{ccccccc} 
u_R & u_R & t_R & U_R & U_R & U_R \\
    \begin{block}{(cccccc) c} 
    0 & 0 & 0 & v & 0 & 0  &\  q_L\\
        0 & 0 & 0 & 0 & v & 0 & \ q_L\\
            0 & 0 & 0 & 0 & 0 & v & \ q_L\\
    \Lambda_u & 0 &  x_1^* & y_1^* & 0 & 0 &\ U_L\\
    0 &     \Lambda_u  &  x_2^* & 0 & y_2^* &  0 & \ U_L\\
     x_1 &  x_2 & \Lambda_t & z_1^* & z_2^* & z_3^* & \ \; T_L\,,
\\
\end{block}
\end{blockarray}
\end{align}
 where $\Lambda_{u,t}$ are invariant mass terms, and 
Yukawa couplings multiplying the VEVs components  
are left understood.
Already this simple construction has some nice features: there are no tree level 
masses for the   `light' fields $q_L\,,u_R\,,t_R$, and  
since by assumption $v \ll x_i,y_i,z_i, \Lambda_{u,t}$   the matrix has a 
see-saw like structure that strongly suppresses light-heavy LH mixings,
ensuring the approximate unitarity of the light quarks mixing matrix.  
The determinant   
$\det(\mathcal{M}) = v^3 \Lambda_u\left[ 
|X|^2 - \Lambda_t \Lambda_u \right]$
  has the correct parametric dependence $\sim v^3 \Lambda^3$ 
to give rise to  three light and three  heavy  eigenstates. 
From  the determinant we can read off which operators in the Yukawa Lagrangian 
must be allowed by $U(1)_F$ 
to ensure $\det(\mathcal{M}) \neq 0$. They are: 
\begin{equation}
\label{eq:yuk6}
        \mathcal{L} \! \supset  \bar q_L H_u U_R   + \Lambda_u \bar U_L u_R + \left\{
    \begin{matrix}      
     \Lambda_t \bar T_L t_R \qquad  {\rm or} \qquad\qquad
    \\
     \bar T_L X u_R +  \bar U_L X^\dagger  t_R.  
     \end {matrix} 
     \right.
\end{equation}
Consider now a generic $U(1)$ that is unbroken by the Yukawa terms in \eqn{eq:yuk6}.
Denoting the $U(1)$ charge of a field 
with the same symbol that denotes the  field, the following  charge relations  must be 
satisfied:\footnote{Whenever no confusion can arise, 
we will use the same symbol to denote a field multiplet and its generic $U(1)$ charge.
We  will instead use respectively  $F_\phi$  and $\mX_\phi$ to denote the 
$U(1)_F$ and $\UPQ$ charges of the field $\phi$.} 
\begin{equation}
\label{eq:conditions6}
 q_L-U_R = H_u, \ U_L-u_R = 0, \   \left\{
 \begin{matrix}  
T_L-t_R=0\qquad  {\rm or} \qquad\quad  \; \\ 
T_L-u_R=  t_R -U_L=X. 
  \end{matrix}
   \right. 
\end{equation}
The generic expression for the coefficient of the mixed $U(1)\times SU(3)_c$ 
anomaly is  
\begin{equation}
    \mathcal{A} = 3 q_L -2 u_R - t_R +2 U_L + T_L - 3 U_R  \,,
\end{equation}
and after imposing the conditions in \eqn{eq:conditions6},  we obtain 
\begin{equation}
    \mathcal{A} =  
    \left\{
    \begin{matrix} 
       3 H_u + (T_L-t_R) = 3 H_u     \quad\quad \  \  {\rm or} \  \qquad \qquad 
        \\
     \;  3 H_u + (T_L-u_R)+(U_L-t_R) =3 H_u  \,. 
       \end{matrix} 
\right.
\end{equation}
Now, under the assumption that the down-quark sector replicates  the same Yukawa structure than the up-sector, 
the total anomaly is $\mathcal{A} = 3(H_u + H_d)$. 
Let us recall  
that for the local $U(1)_F$ we need 
$\mathcal{A}_F=0$, 
while for $\UPQ$ we need $\mathcal{A}_{\rm PQ}\neq 0$. This requires two Higgs doublets
($H_u\neq \tilde H_d$) with  
opposite $F$-charges $F_{H_u} = - F_{H_d}$.  
The  bilinear term $H_u H_d$ would then be permitted by $SU(2)_L\times U(1)_Y \times U(1)_F$ invariance,  however, since  $\mX_{H_u} + \mX_{H_d}\neq 0$,   it would badly break the PQ symmetry at $D=2$. 
Thus this simple model, although presenting some interesting features,  does not allow to establish 
an axion-flavour connection. Nevertheless, this example 
illustrates rather clearly which   strategy  
should be followed  to  study the viability of other constructions.\\ [-8pt]

\subsection{Strategy for model building}
\label{sec:strategy}

Models of flavour in which a local gauge symmetry can  
automatically yield a high-quality QCD-axion 
can be constructed by implementing the following steps: \\ [-8pt]

\noindent 
(1) {\it Mass matrix determinant and anomaly conditions.}
Given a set of scalar and fermion  multiplets 
transforming under a  semi-simple  flavour factor $\GF$,   
 write down the most general $\GF$ invariant Yukawa potential, and  
verify that the leading terms of  the determinant of the fermion mass matrix  have the 
parametric structure  $\sim  v^3 \Lambda^{n_Q-3} + \dots $ 
where the ellipsis represent possible terms of $O(v^4)$ or higher, while terms   $v^j \Lambda^{n_Q- j} $ 
with $0\leq j \leq 2$, if present, must eventually be  forbidden by the  $U(1)_F$ symmetry. 
Next, for each choice of the Yukawa terms that can yield the correct parametric dependence,  
find  which accidental $U(1)$'s arise,  and verify that  $U(1)$  charge relations 
allow for solutions of the anomaly coefficient conditions  $\mathcal{A}_F =0 $ and  
$\mathcal{A}_{\rm PQ}\neq 0$.  \\ [-8pt]

\noindent
 (2) {\it Identifying $U(1)_F$ and $\UPQ$.}
The  gauge invariant kinetic term 
for the set of $n = n_\psi + n_\phi$  fermion and scalar multiplets 
enjoys a $U(1)^n$  rephasing symmetry. 
Each Yukawa operator and each non-Hermitian scalar operator, when  allowed,   
imposes one condition on the otherwise arbitrary phase redefinitions, 
and reduces  the number of the $U(1)$ symmetries by one unit. 
We aim  to find a set of $n-4$ Yukawa and scalar operators that  
will reduce the $U(1)^n$  symmetry 
down to $ U(1)_\mY\times  U(1)_\psi \times  U(1)_{\rm PQ}\times U(1)_F $.
 The first factor is  hypercharge, for which we have $\mY_{H_d} = - \mY_{H_u}=\frac{1}{2}$
(in the case of a single  Higgs doublet   $H_d = \tilde H_u$), while all the exotic scalars have $\mY=0$. 
The second factor is fermion number, under which all the fermion multiplets  have 
the same charge while all the scalars are neutral. Note that 
when only the quark sector is considered, $U(1)_\psi$ can be  simply identified 
with baryon number.  $U(1)_\psi$   is automatically preserved 
because Lorentz invariant fermion operators of the Majorana type   
cannot appear in the quark sector.
The third factor is the sought PQ symmetry. Note that the $U(1)_\psi$ and $\UPQ$ charges,  
being global, can always be redefined  by means of  a shift proportional to hypercharge or $F$-charge.
%
 %
Now, requiring that a specific set of  $n-4$  Yukawa and scalar 
operators are allowed by  $U(1)_{\mathcal{Y}}$ and by a certain $U(1)_F$  local symmetry, provides $n-4$ constraints.   
It follows that  the charges of all the fermions ($\Psi_f$)   can be expressed 
in terms of linear combinations of four reference charges. The charges of the Higgs doublets,  that do not carry fermion number, depend on three reference charges,  while for the SM singlet scalars ($\Phi_s$) two charges suffice, since they do not carry 
neither   fermion number nor hypercharge. 
We will choose the charge $\zeta$ of one fermion multiplet
to represent fermion number and the charge of one SM singlet  scalar $\varphi$, to represent   
one combination of the $F$- and PQ charges. For the remaining two reference charges we 
define  $h_+ = H_d + H_u$ and $h_- = H_d - H_u$. Note that for $h_+$ the hypercharge contribution cancels, 
so that $h_+$  can be taken as a second combination of  $F$- and PQ. 
Note also that since  $h_- $ contains a hypercharge contribution, 
it cannot appear in the charge combinations $\Phi_s$.\footnote{In models with a single Higgs doublet
$H$ one should replace $h_+$  with the charge of a second hyperchargeless  scalar, and $h_- \to  H$. } 
All in all we can write:
\begin{eqnarray}
\label{eq:fcharges}
    \Psi_f &=& a_f \varphi + b_f  h_+  +  c_f h_- + d_f \zeta \,, \\
\label{eq:scharges}
   \Phi_s   &=& a_s \varphi + b_s h_+    \,,\\ 
\label{eq:sanomaly}
\mathcal{A}  &=&   a\, \varphi + b\, h_+  = 
\mA_F + \mA_{\rm PQ}\,, 
      \end{eqnarray}
 where the coefficients  $a_{f,s},b_{f,s},c_{f},d_f$ and $a,b$,
 are determined by the   $n-4$ conditions 
 on the charges. The anomaly coefficient in the last expression 
 does not depend on $\zeta$  and $h_-$ since fermion number and hypercharge have  
 no  QCD anomaly while, as it is put in evidence in the second equality, it 
 can receive contributions from the other two $U(1)$'s. 
If the coefficients $c_f$ are all proportional to the corresponding hypercharges $y_f$,
 then  $h_-$   can be identified with the hypercharge generator.  
To  achieve this   a suitable shift of the global charge $\zeta$  might be needed,
 e.g.  $\zeta \to \psi - \alpha\, h_-$ 
 in such a way that  for all fields  $c_f- \alpha\, d_f=y_f$. 
 Finally,  the $U(1)_F$   charges 
 are identified by the condition 
$\mathcal{A}_F= 0$,  
which yields $f \propto a\, h_+ - b\,\varphi$ and $\chi \propto b\, h_+ + a\, \varphi$, where 
$f$ and $\chi$ are  normalization factors for the $U(1)_F$ and $\UPQ$ generators.

One should next verify that  PQ breaking operators  $O_\subPQv$  
will only appear at a sufficiently high operator dimension. The search for such operators 
is  greatly simplified by the fact that their overall charge must be proportional to the 
anomaly coefficient $\mathcal{A}$,  since this is the necessary condition 
for $O_\subPQv$   being allowed by the gauge symmetry  while carrying an 
overall PQ charge.  \\ [-8pt]


\noindent
 (3)  {\it Identifying the physical axion}.  
As  we have mentioned above, the PQ charges are not uniquely defined  since any combination 
of the accidental $U(1)$ and of the diagonal generators associated with the  gauge symmetries 
defines a new global symmetry~\cite{Kim:1981cr,Ernst:2018bib,DiLuzio:2020qio}. 
The {\it physical}  axion must be, however,  a massless state orthogonal to the Goldstone bosons of the 
broken flavour 
symmetries. This implies the following conditions: 
\begin{equation}
\label{eq:ortho}
\sum_S \vev{S^\dagger T_C S } \mX_S = \sum_S \vev{S^\dagger S} F_S \mX_S = 0\,,  
\end{equation}
where $S$ represents any  scalar field 
and $T_C$ are the Cartan generators of the non-Abelian subgroups. 
These conditions can be satisfied via a redefinition of the PQ charges,   
that will then identify the 
particular $\UPQ$ whose Goldstone boson is 
physical axion.\footnote{If the model contains two or more 
PQ charged Higgs doublets, upon EW symmetry 
breaking their charges must be further redefined  to satisfy 
$\sum_i \mathcal{Y}_{H_i} \mX_{H_i} v^2_i  = 0$, to ensure that the axion 
does not mix with the  Goldstone boson of   $U(1)_{\mathcal{Y}}$~\cite{Zhitnitsky:1980tq,Dine:1981rt}.}
Let us note at this point that it is mandatory that at least one 
among the  scalars with   VEVs  $V\gg v$ will have 
 the coefficient   $b_s$ of its charge $\Phi_s$  (see \eqn{eq:scharges}) 
 different from zero,  otherwise  one global $U(1)_{h_+}$ would remain unbroken
 below the scale $V$. 
 The surviving symmetry  will eventually get broken at the EW scale,    
however, the resulting axion would then lie  
in the Higgs direction. 
Formally, this occurs because the orthogonality conditions \eqn{eq:ortho} imply that 
after redefinition, the   PQ charges of the Higgs doublets (and consequently  also the charges of the SM fermions
coupled to the Higgs)  become extremely large, of $O(V^2/v^2)$. It can be shown that  
 the axion would then couple to the  SM fields with an effective suppression scale of  $ O(1/v)$ so that, 
 much alike the  Weinberg-Wilczek axion~\cite{Weinberg:1977ma,Wilczek:1977pj}, 
it would be phenomenologically ruled out.  \\ [-8pt]

%
%

\noindent
 (4) {\it Flavour structure.}
After constructing the axion sector according to the previous steps, 
it remains to be seen which flavour structure can arise from the model.  
To carry out this study  one should write down the most general scalar potential  in terms 
of the complete set of $\GF\times U(1)_F$  invariant scalar operators, which  will be also  
PQ conserving by construction. Minimisation of the potential will then yield 
a  fermion mass matrix,  of which \eqn{eq:66Matrix} is one example. The goal is to obtain 
the correct hierarchies among the light quark masses without the need  of  
strong hierarchies among the parameters of the scalar potential or among the 
couplings  of the Yukawa operators. Previous studies carried out within a  model based on the  
flavour symmetry  $SU(3)_Q\times SU(3)_u\times SU(3)_d$  (but with no axion) have proven to be 
successful in this endevour~\cite{Nardi:2011st,Espinosa:2012uu,Fong:2013dnk},  and indicate that 
there should be no fundamental impediment to achieve the same result with   $\GF\times U(1)_F$. 
However, in the present case carrying out such a  study  is remarkably more complicated. This is because 
since we need to keep trace of the PQ anomaly,  we have to work with 
the complete renormalizable UV theory, 
rather than much more simply with an effective theory involving only the  SM quarks, 
as was  done e.g. in Refs.~\cite{Nardi:2011st,Espinosa:2012uu,Fong:2013dnk}. 
We  anticipate that this step requires a computationally intense analysis, since it involves 
multiple minimisations of a complicated multi-fields scalar potential to find the structure of the VEVs, 
nested with  minimisations with respect to the couplings of the  scalar and Yukawa  potentials to find 
the set of parameters that can reproduce the SM quark flavour structure. 
Proper minimisation  of the scalar potential  
requires a careful  parametrisation of the scalar multiplets, so that they will contain 
only  physical degrees of freedom,  with all  redundant components 
that can be gauged away removed.  
In the Appendix we derive explicit expressions for the field multiplets 
containing only   non-redundant degrees of freedom. 
This parametrisation  has been used in the numerical minimisation. 



 \section{A viable $\mathbf{SU(3)\times SU(2) \times U(1)_{F}}$  model} 
\label{sec:model}

%
%
%

\noindent
As a concrete application, let us describe a model in which 
 $\UPQ$  arises automatically and is  protected up to $D=10$, while strong mass hierarchies arise 
 upon  minimisation of the scalar potential of a set of EW singlet scalars.
 The flavour group is the same  
 $SU(3)\times SU(2) \times U(1)_{F}$ of the  model discussed  in Section \ref{sec:modelbuilding},  
but we extend the fermion content 
of  Eq.~(\ref{eq:fermions})
by including a pair of  flavour singlets EW  doublets $Q_L=(t,b)_L^T$ and $Q_R=(T,B)_R^T$. 
The up-type quark sector thus contains the following fields:  
\begin{eqnarray}
\nonumber
    &q_L\sim (\bold{3},\bold{1}), \  u_R\sim (\bold{1},\bold{2}), \  t_R \sim (\bold{1},\bold{1}), \   Q_L \sim (\bold{1},\bold{1}),\quad \quad \\
    &U_R \sim (\bold{3},\bold{1}), \,  U_L \sim (\bold{1},\bold{2}), \, T_L \sim (\bold{1},\bold{1}),  \,  Q_R \sim (\bold{1},\bold{1}).\quad\quad
\label{eq:fermcont}
\end{eqnarray}
Together with the down-type singlets $d_R, b_R, $ $D_R, D_L, B_L$,  this set of fermions  is manifestly free from  QCD and $\GF$ gauge anomalies.  
Let us also assume  the same scalar sector given in Eq.~(\ref{eq:scalars}). 
The most general Lagrangian invariant under $\GF = SU(3)\times SU(2)$ yields the following mass matrix: 
%
%
\begin{align}
\label{eq:77Matrix}
\mathcal{M}_u=\quad 
\begin{blockarray}{cccccccc} 
 u_R & u_R & t_R & U_R & U_R & U_R & Q_R\\
    \begin{block}{(ccccccc) c} 
 0 & 0 & 0 &  v & 0 & 0    &  z_1 &\  q_L \\
  0 & 0 & 0 & 0 &  v & 0   &  z_2 & \ q_L  \\
  0 & 0 & 0 & 0 & 0 &  v   &  z_3  & \ q_L \\
 0 & 0 & v& 0 & 0 & 0    & M  &\  Q_L \\
 \Lambda_u & 0 & \, x_1^* &  y_1^* & 0 & 0 &  0 & \ U_L \\
 0 &     \Lambda_u  & \, x_2^* & 0 &  y_2^* &  0 & 0 & \ U_L \\
 \, x_1 & \, x_2 & \Lambda_t &  z_1^* &  z_2^* &  z_3^* &  v &\  \; T_L\,,
\\
\end{block}
\end{blockarray}
\end{align}
where all Yukawa couplings 
are left understood. 
The determinant of this mass matrix has the required parametric structure $\det \mM_u \sim O(v^3\Lambda^4) + O(v^5\Lambda^2)$. 
Each entry in $\mM_u$ corresponds to one Yukawa operator allowed by $\GF$, but not necessarily by $U(1)_F$. 
For each operator that we require to be also $U(1)_F$ invariant, 
we get  one condition on the Abelian charges.  
Let us see how many conditions  should  arise from the Yukawa sector.
In \eqn{eq:fermcont} we have three fermion doublets plus five singlets. Taking into account the
additional five down-type $SU(2)_L$  singlets we have a total of thirteen fermion multiplets.
We should impose no more than twelve linearly independent 
conditions on the charges of the Yukawa operators, 
in order to allow for an unbroken  $U(1)_\psi$
acting on the fermions.\footnote{In general, a certain number of additional  Yukawa operators 
corresponding to conditions that are linearly dependent from the ones imposed will also be allowed by accident.}
We should also impose no less then twelve conditions to avoid two independent $U(1)_\psi \times U(1)_{\psi'}$, 
which would  imply the existence of a new `baryon number' that could stabilize some exotic quarks,  giving  rise to dangerous strongly interacting long lived 
relics~\cite{Nardi:1990ku,Jacoby:2007nw,Kusakabe:2011hk}. The following operators are mandatory: 
\begin{equation}
 \label{eq:detConditions}
[H_u \bar{q}_L^\alpha\, U_{R, \alpha}\,,\  \ 
 \Lambda_u  \bar{U}^i_L u_{R,i} \,,\ \ 
 H_u \bar{Q}_L t_R]\,, \ \  
 \bar{q}_L^\alpha Z_\alpha Q_R \,, 
    \end{equation} 
where for the sake of clarity we have written explicitly  the flavour indices 
(Latin indices $i,j,\dots $ refer to $SU(2)$, Greek indices $\alpha,\beta,\dots$ to $SU(3)$). 
The first three operators in the square brackets 
are specific of the up sector, and should be mirrored in the down-quark sector, 
($U_{L,R} \to D_{L,R}$, $u_R\to d_R$, $t_R\to b_R$,   
$H_u \to H_d$) providing three additional conditions. 
The fourth one 
already involves both the up- and down-type quark doublets. 
The first two operators in \eqn{eq:detConditions}  are necessary to ensure that $\det \mM_u\neq 0$. 
The third one  gives rise to a mass term of $O(v)$ at tree level. This 
accounts for the fact that the mass of one up-type quark (the top) is not suppressed with respect to the EW  scale, 
and hence it should not originate from an effective operator. The presence of this operator, together with the previous two,   also suffice to ensure a non-vanishing determinant. 
The last operator is needed to ensure  that only three 
up-type quarks will  be present below the EW scale. In fact  it  removes 
one linear combination of the three $q_L$   from the low energy spectrum by
providing a mass $O(\Lambda)$.  

The operators in \eqn{eq:detConditions}  do not involve  the quark $T_L$, 
nor the  $X,Y$ scalars that  would not contribute to dynamical generation of the 
mass hierarchies. 
This is remedied by requiring the following additional operators 
 \begin{eqnarray}
 \label{eq:Foperators}
[  \bar U_L Y^\dagger U_R, &&  
 \ \  \  \bar T_L X u_R], \ \ \  \ \ \ \  \Lambda_t \bar T_L t_R. 
  \end{eqnarray}
The first two in the square brackets  
suffice to have a $X,Y$-dependent Yukawa Lagrangian and to couple $T_L$.
They are replicated in the down-quark sector ($U_{L,R}\to D_{L,R}$, $T_L\to B_L$,  $ u_R\to d_R$). The last 
one completes the required list of twelve operators yielding linearly 
independent charge conditions (seven from \eqn{eq:detConditions} and five from \eqn{eq:Foperators}). 
Although the presence of the corresponding down-type operator  $\Lambda_b \bar B_L b_R$ is   not imposed, one can verify that it  arises by accident.

Let us now assume that there is only one Higgs doublet $(H_u = \tilde H_d$). Together with 
$X,Y,Z$ and the thirteen fermion multiplets, there are seventeen fields. Then, on top of the twelve 
Yukawa conditions, we need to impose 
one additional condition in the scalar sector to reduce $U(1)^{17}\to U(1)^4$.
The set of scalar fields in \eqn{eq:scalars} does not allow to write a non-Hermitian operator 
involving the Higgs doublet, the only possibilities  
(with obvious contraction of the $ SU(3)\times SU(2)$ indices)  are $XY^\dagger Z,\, XYZ^\dagger,\, 
ZYY,\, X(YY^\dagger) X$. By choosing to allow in turn each one of these operators, and 
solving the set of thirteen linearly independent constraint on the charges, 
one can verify that   the first operator yields a vanishing overall anomaly $\mA=0$, and thus 
there is no PQ symmetry.  For each one of the other three possibilities, one 
obtains that at least one of the three remaining operators acquires an overall charge proportional 
to the anomaly coefficient. Thus, after  imposing the condition $\mA_F=0$, this operator will be allowed
by $U(1)_F$; however, since it will have to carry a non-vanishing PQ charge $\propto \mA_{\rm PQ}\neq 0$, 
it will break $\UPQ$ at the renormalizable level. We can  conclude that with a single Higgs doublet, 
the model does not allow to establish an axion-flavour connection. 

In order to proceed, let us then assume two Higgs doublets $H_u, H_d$. 
To reduce the grand total of eighteen  rephasing symmetries to $U(1)^4$ 
we now need to impose  {\it two}  additional conditions 
on the scalar couplings.  Even with two Higgs doublets the set of fields in \eqn{eq:scalars}
still   does not allow to write a non-Hermitian operator involving the Higgs fields.
This implies that the two scalar conditions will involve only $X,Y,Z$. As a consequence 
the charges of these fields will depend on a single charge $\varphi$ (i.e. in \eqn{eq:scharges} $b_s=0$).  
 As explained in point (3) of the previous section,  this implies that an  accidental 
$U(1)$ symmetry  remains unbroken by the large VEVs, resulting in an axion that mainly sits in the Higgs doublets, and that couples too strongly to the SM  fields. 

A simple way to avoid this issue is to introduce a non-Hermitian coupling between the Higgs doublets 
and the EW singlet scalars, that will ensure $b_s\neq 0$. This can be easily done with a further 
enlargement of the field content. 
We choose to add to the set in \eqn{eq:scalars} a  scalar multiplet $K_\alpha\sim (\bold{3},\bold{1})$ transforming like $Z_\alpha$.
We can then write the operator $(Z^\dagger K)(H_uH_d)$  (see \eqn{eq:Soperators} below)
which forces the charges of the heavy  scalars to acquire a dependence on $h_+$. 
We couple this new scalar to the Yukawa sector by replacing one of the  previous Yukawa operators  
with a new one involving $K$, more precisely   $\bar B_L X d_R \to \bar B_L K^\dagger D_R$, 
The five conditions corresponding to \eqn{eq:Foperators} now are:
 \begin{eqnarray}
 \label{eq:FoperatorsK}
 \nonumber
  \bar U_L Y^\dagger U_R, &&   \bar T_L X u_R, \ \  \ 
  \Lambda_t \bar T_L t_R,  \\
  \bar D_L Y^\dagger D_R, &&  \bar B_L K^\dagger D_R\,. 
  \end{eqnarray}
After imposing the corresponding charge conditions,  
two additional Yukawa operators $\bar{U}_L X^\dagger t_R$ and 
$\bar{D}_L X^\dagger b_R$ are also  allowed by accident. 
Note that the set of Yukawa operators is not completely symmetrical between  the up and down sectors. 
$  \Lambda_t \bar T_L t_R$ and  $\bar{T}_L X u_R$  are not mirrored in the down sector, which instead includes 
$ \bar B_L K^\dagger D_R$  that has no counterpart in the up sector.

We now need to impose 
three conditions among the scalars. We require $U(1)_F$ invariance of the following  operators: 
\begin{eqnarray}
&&Z^{\dagger\alpha} K_\alpha H_u H_d,\ \ \epsilon^{\alpha\beta\gamma} 
X^\dagger_i Y^i_\alpha Z_\beta K_\gamma, \ \
\epsilon_{ij}X^i Y^j_\alpha Y^{\dagger\alpha}_k X^k ,  \qquad
\label{eq:Soperators} 
\end{eqnarray}
Once the corresponding charge conditions 
are imposed, 
the additional operator  $\epsilon_{ij}\epsilon^{\alpha\beta\gamma}X^i Y^j_\alpha K_\beta Z_\gamma$
is also accidentally allowed. This completes the list of the allowed Yukawa and scalar operators 
of the model.

As for the PQ origin and quality issues, we have verified that 
(with natural values of the charges) $U(1)_F$ forbids all 
renormalizable PQ breaking operators,  so that indeed the origin 
of $\UPQ$  is accidental.  Moreover, the first PQ breaking 
operators arise at $D=11$, so that $\UPQ$ is sufficiently 
protected, and the quality issue solved.\footnote{The $D=11$ 
PQ breaking operators 
contain the $\GF$ invariant  factor  $(K Z^\dagger)^4$ joined to one 
of the three   $\GF$ invariant strings 
$KYY$, $Z^\dagger Y X$ or $Z^\dagger Y X^\dagger$.  Each power of  
$(K Z^\dagger)$ can also be replaced by 
$(H_u H_d)$ since it  carries the same $U(1)_F$ and PQ charges. 
However,  the effects of these additional operators are suppressed 
by powers of $v^2/V^2$ and are thus irrelevant.} 

Carrying out the  minimisation of the full scalar potential, 
inserting the resulting VEVs in the mass matrices $\mM_{u,d}$ and
extracting the light $O(v)$ eigenvalues, 
yields the quark masses displayed in 
Table~\ref{table:masses}.\footnote{In addition to 
the non-Hermitian scalar operators discussed above, 
the  scalar potential also contains many additional 
Hermitian operators. Being $U(1)_F$ neutral they play no role 
in determining $U(1)_F$ and $\UPQ$, and they are generally  
also `flavour irrelevant' in the sense of Ref.~\cite{Fong:2013dnk}. 
This happens  when an operator is accidentally invariant under a   
symmetry larger than $\GF$ (e.g.  $Z^{\alpha \dagger} Z_\alpha$
that carries an $O(6)$ symmetry),  
so that its contribution to the potential energy does not  
depend on the particular configuration of the VEVs 
components. (Details on the numerical minimisation  
procedure can be provided upon request.)}
Notice that the 
reference values of the quark masses are evolved (in the  $\overline{\rm MS}$ scheme) up to $10^8 \,$GeV~\cite{Huang:2020hdv}.

\begin{table}[t!]
\begin{center}
\begin{tabular}{l c c } 
\hline
\rule{0pt}{10pt}  & \text{model} \quad & \ \text{experimental}\\[0.2em]    
\hline
 \rule{0pt}{12pt}$m_b (\text{GeV})$\ \  & \ 1.5 \ & \ 1.5\  \\[0.3em]  
 $m_c (\text{GeV})$ & 0.5 & 0.4 \\[0.3em] 
  $m_s (\text{MeV})$ &\ 20 &\  30\\[0.3em] 
 $m_d (\text{MeV})$ & 0.5 & 1.5 \\[0.3em] 
   $m_u (\text{MeV})$ & 0.3 & 0.7 \\[0.3em] 
\hline
\end{tabular}
\caption{Quarks mass values obtained in the reference model (see text) and their experimental values evolved at the scale $\Lambda = 10^8$GeV. 
A top mass value $m_t(\Lambda) = 102.5\,$GeV~\cite{Huang:2020hdv} has been used to fix the overall fermion mass scale.
}
\label{table:masses}
\end{center}
\end{table}

Although in our  model we set  the  scale where the first breaking of the  flavour symmetry occurs at  $10^{11}\,$GeV, 
the complete breaking proceeds in various steps, separated  by hierarchical  energy gaps. The  reference scale of 
$10^8\,$GeV  is a rough average between the highest and the lowest  breaking scales, and thus representative of  the 
scale  where  the quark Yukawa couplings are generated.
The comparison between the results of the minimisation and the values of the running masses 
in Table~\ref{table:masses} is thus affected by  various uncertainties, and should be taken with a grain of salt.  The  important point is that  in the numerical study  the ratios between the dimensionless Lagrangian couplings  were constrained to never exceed a 
factor of $\mathcal{O}(10)$, 
so that the  hierarchical pattern arises dynamically. 
In fact, 
starting  from non-hierarchical values of the fundamental parameters,  
we obtain for the VEVs components 
$|x_1|,\, |y_{1,2}|,\, |z_3|,\, |k_2| \sim O(\Lambda)$,  
$|x_2|,|z_{1,2}|\sim O(10^{-5} \Lambda)$,   
$|k_{1,3}| \sim O(10^{-6} \Lambda)$ with $\Lambda = 10^{11}\,$GeV.   
We also find remarkable that the same set of VEVs allows to reproduce both 
the stronger mass hierarchy of the up sector, and the milder hierarchy of the 
down sector, which in terms of a small parameter
$\epsilon\sim 0.2$  (of the order of  the Cabibbo angle) can be written as: 
\begin{equation}
    \begin{split}
        m_u:m_c:m_t=\epsilon^7:\epsilon^3:1 \ \ \\
        m_d:m_s:m_b=\epsilon^6:\epsilon^4:\epsilon^2\,. 
    \end{split}
    \label{eq:massratios}
\end{equation}
This is possible thanks to the asymmetry  between the  operators contributing to 
the mass matrices in the up and down sectors mentioned above. 

We finally note that although the model contains two Higgs doublets, it is 
fundamentally different from well known 2 Higgs Doublet Models (2HDM) obtained by 
enlarging the SM Higgs sector. In first place the scalar potential of 
2HDM does not carry a PQ symmetry, and hence all the scalars are massive. 
Here, instead, the usual PQ breaking operators  $m^2 H_u H_d$  and $\lambda_5 (H_u H_d)^2$
are forbidden by the $U(1)_F$ symmetry, and a massless axion then arises. 
$U(1)_F$ also ensures, without the need of additional assumptions, that the Yukawa sector 
features natural flavour conservation~\cite{Glashow:1976nt}.  
However, as in 2HDM, there are additional massive Higgs scalars that 
must be sufficiently heavy to have so far escaped detection. 
These results allow us to conclude   that an axion-flavour connection can be a realistic \textit{ansatz} to tackle 
the strong CP problem and the  puzzle  of the flavour mass hierarchies in one fell swoop.
On another note, it is clear that  the different structure of $\mathcal{M}_u$  and $\mathcal{M}_d$ 
implies that their diagonalisation will involve structurally different  bi-unitary rotations $\mV^{u,d}_{L,R}$, and this
might yield  a  too large relative misalignment between the up and down LH mixing matrices.  
Indeed, by denoting as $V_L^{u,d}$ the restriction of  $\mV^{u,d}_{L}$ to the subspace of the 
three light eigenvalues,  we find that in our model  $V_{\mathrm{CKM}} =  V_L^u V_L^{d\dagger}$  
does not give a satisfactory description of the experimental  quark mixing matrix.\footnote{$V_{\mathrm{CKM} }$ is also non-unitary. However, deviations from unitarity are of $O(v/V)$ and thus negligible.}  
This  hints to the fact that the flavour symmetry on which  we have based our model might be  too simple. Since 
$\GF$  does not distinguish between the two sectors (only $U(1)_F$ does),  a  single set of VEVs
contributes to both the mass matrices, and this  seems too constraining to account 
for the two different hierarchical patterns in \eqn{eq:massratios} while keeping at the same time a sufficient
similarity between the matrix structures $\mM_{d} \sim \mM_u$.   
For example, splitting the  $SU(2)$ factor in $\GF$ according to   $SU(3)\times SU(2) \to 
SU(3)\times  SU(2)_u\times SU(2)_d $ would imply doubling the scalar fields carrying $SU(2)$ indices  
 $X, Y \to X_u, Y_u, \, X_d, Y_d$, and this can provide a way to circumvent  the hierarchies/mixings clash~\cite{inprep}. 


 \section{Outlook and conclusions}
\label{sec:outlook}
The   {\it origin} and  {\it quality}  problems of the PQ symmetry  can be  solved by 
postulating a  new local symmetry,  suitable to enforce automatically the invariance 
 of the renormalizable Lagrangian under a global anomalous $U(1)$ symmetry,    
and able to protect this symmetry from explicit breaking by effective operators up to some 
remarkably large dimension. 
We have shown that  a class of   symmetries that can yield  this result 
are based on  a semi-simple {\it rectangular} gauge group  $\GF$ 
extended by a local $U(1)_F$ factor.  Since neither group factors  of large degree nor large fermion representations are required,  such symmetries 
can be straightforwardly interpreted as flavour symmetries acting on the SM quarks.  
The VEVs of the scalar multiplets that break the flavour symmetry 
also break $\UPQ$ giving rise to an axion, while  the 
structure of these VEVs can be responsible for   the SM flavour hierarchies
that are generated dynamically, without the need of hierarchical fundamental parameters.  
We have provided  one example in which the SM mass ratios  are reproduced as a result of 
minimisation of the scalar potential. Clearly, other models can be constructed along the same lines 
outlined in this paper, and we are currently  exploring the landscape of possible models 
in search for the most economical and elegant realisations. 
From the phenomenological point of view, it is interesting to note that while  the flavour  (and PQ) breaking 
scale must be sufficiently large to render   the axion invisible $f_a \sim V \gg v$, 
experimental signatures of this type of constructions 
can still be within experimental reach.  This is because small mass ratios like  $m_u/m_t \sim 10^{-5} $  
must originate from a hierarchy between the  VEVs of   some   scalar multiplets components. 
Then, even if we assume that the largest VEVs are of the order of the axion scale, e.g.  $f_a \sim 10^{10}\, $GeV, 
some scalar component must  acquire a much smaller VEV,  of  the order of  $f_a m_u/m_t\lesssim 100\,$TeV. 
Hence,  it is not unreasonable to speculate  that some flavoured gauge bosons  could have a mass  below 
the 100 TeV scale,  and could  then give rise to observable effects in high precision flavour-related measurement.  
Clearly, after a VEV structure that is able to reproduce the SM flavour parameters has been identified, 
it would be  not difficult  to reconstruct, from the values of the VEVs  of the scalar components,
the spectrum of the flavoured gauge bosons, and derive phenomenological predictions. This type 
of phenomenological studies is left for future work. 

\section*{Acknowledgments}
We acknowledge several discussions with F.P. Di Meglio, G. Grilli di Cortona, 
L. di Luzio and A. Salvio.  L.D. is  supported by the European Union’s Horizon 2020 research and innovation programme under the Marie Skłodowska-Curie grant agreement No 101028626 from 01.09.2021.
E.N. is supported by the INFN Iniziativa Specifica Theoretical Astroparticle Physics (TAsP-LNF). 
C.S. acknowledges hospitality and partial financial support from the LNF theory group.

 \newpage

  \onecolumngrid

\section*{Appendix: 
Parametrization of the scalar fields}
 \label{sec:parametrization}
 
 In this Appendix we describe the parametrization for a set of scalar multiplets transforming under  
 a gauge symmetry $SU(3)\times SU(2)\times U(1)$. 
 We  first write the matrices of fields in their general singular value decomposition (SVD), next we 
 manipulate the resulting expressions so that they will provide a faithful parametrization of the scalar multiplets
 by eliminating the redundant degrees of freedom (d.o.f.). Finally we make use of the full local gauge symmetry 
 to gauge away additional d.o.f.   This last step fixes completely the gauge. 
 We carry out this study for the set of scalar representations corresponding to the multiplets $Y,\,Z,\,K,\,X$ introduced in the model of 
 Section~\ref{sec:model}, however,  this example is  sufficiently general to render clear how one should proceed  with other sets of multiplets.

The SVD gives the following forms for the scalar fields (recall that in terms of the number of rows and columns 
the field matrices are
 $Y=Y_{3\times 2},\, 
 Z=Z_{3\times 1},\, 
 K=K_{3\times 1},\, 
 X=X_{1\times 2}$):  
\begin{eqnarray}
\label{eq:Y}
Y &=& \mV_Y^\dagger \hat Y e^{i\xi_Y} \mU_Y,  \quad 
[6_R + 6_I \ \mathrm{vs.} \  6_R+8_I], 
\quad
\mV_Y \in SU(3), \ \ \mU_Y \in SU(2), \  \ \hat Y = 
\begin{pmatrix}
y_1 & 0 \cr
0 & y_2  \cr
0 & 0& 
\end{pmatrix}, 
\\
\label{eq:Z}
 Z &=& \mV_Z^\dagger \hat Z e^{i\xi_Z},  \  \qquad [3_R+3_I\  \mathrm{vs.} \ 4_R+6_I], \quad \mV_Z \in SU(3)  \qquad \hat Z = 
\begin{pmatrix}
 z_1 \cr
0    \cr
0
\end{pmatrix},
 \\
 \label{eq:K}
  K &=& \mV_K^\dagger \hat K e^{i\xi_K},   \qquad [3_R+3_I\  \mathrm{vs.} \ 4_R+6_I], \quad \mV_K \in SU(3)  \qquad \hat K = 
\begin{pmatrix}
 k_1 \cr
0    \cr
0
\end{pmatrix},
 \\
 \label{eq:X}
 X &=& \hat X e^{i\xi_X} \mU_X,
  \qquad [2_R+2_I\  \mathrm{vs.} \ 2_R+3_I], 
 \quad \mU_X \in SU(2) \qquad \hat X = 
\begin{pmatrix}
x_1 & 0 
\end{pmatrix}.
\end{eqnarray}
In the square brackets we have indicated the number of real (R) and imaginary (I) d.o.f. of the multiplet, 
versus the number of parameters entering their SVD.  The number of redundant 
parameters in the SVD  is: $(+2_I)$ for $Y$,  $(+1_R,+3_I)$ for $Z$ and $K$ and   $(+1_I)$ for $X$.  
The redundancies can be eliminated by constraining the  matrices $\mV$ and $\mU$ to 
depend on a reduced number of parameters with respect to  general special unitary matrices
(3 real angles and 5 phases for $SU(3)$, 1 real angle and 2 phases for $SU(2)$). 
 The matrices $\mU$ and $\mV$ can be thought as describing  general $SU(2)$ and $SU(3)$ 
 finite transformations, and can be parametrized in the following way~\cite{murnaghan1962unitary}:

\begin{equation}
\label{eq:UV}
\mU =  \Theta U, \qquad \qquad
\mV =\Phi V_x V_z V_y,  \quad 
\end{equation}
with
\begin{eqnarray}
\label{eq:UV2}
&& U=\begin{pmatrix}
c_{\alpha_u} & -s_{\alpha_u} e^{-i\beta_u} \cr
 s_{\alpha_u} e^{i\beta_u} & c_{\alpha_u}  
 \end{pmatrix}, \qquad  
\Theta =
\begin{pmatrix}
e^{i\theta} & 0\cr
0 & e^{-i\theta} 
\end{pmatrix} ,  \quad 
\left\{
\begin{matrix}
-\pi \leq \alpha_u < \pi \cr
-\frac{\pi}{2} \leq \beta_u,\theta \leq \frac{\pi}{2} 
\end{matrix}
\right. 
 \end{eqnarray}

 \begin{eqnarray}
\nonumber
&& V_x=\begin{pmatrix}
1&0&0 \cr 
0 & c_{\alpha_x} & -s_{\alpha_x} e^{-i\beta_x} \cr
 0 & s_{\alpha_x} e^{i\beta_x} & c_{\alpha_x} 
 \end{pmatrix}, 
\quad
V_y=\begin{pmatrix}
 c_{\alpha_y} & 0 & -s_{\alpha_y} e^{-i\beta_y} \cr
0&1&0 \cr 
  s_{\alpha_y} e^{i\beta_y} & 0 & c_{\alpha_y} 
 \end{pmatrix} ,  \\  [10pt]
&&   \ V_z=\begin{pmatrix}
 c_{\gamma_z} & -s_{\gamma_z} e^{-i\beta_z} &0 \cr
  s_{\gamma_z} e^{i\beta_z} & c_{\gamma_z} & 0 \cr
0&0&1 \cr 
 \end{pmatrix},  
\quad
 \Phi=
\begin{pmatrix}
e^{i\varphi_1} & 0 &0\cr
0 & e^{i\varphi_2} &0 \cr
0 & 0 & e^{-i(\varphi_1+\varphi_2)}  
\end{pmatrix},   
\quad
\left\{
\begin{matrix}
-\pi \leq \alpha_x, \alpha_y < \pi \cr
-\frac{\pi}{2} \leq \gamma_z, \beta_{x,y,z},  \varphi_{1,2} \leq \frac{\pi}{2} 
\end{matrix}
\right. \,.\quad
\end{eqnarray}
One should pay attention to the ordering in $\mV$ $\sim (V_x\cdot V_z\cdot V_y)$ and to the fact that 
the angle $\gamma_z \in V_z$ is a ``latitude" angle  that ranges in the interval $\gamma_z \in [-\frac{\pi}{2},\frac{\pi}{2}]$ 
while  $\alpha_{x}$, $\alpha_y \in V_{x,y}$ are ``longitude" angle whose range is $   [-\pi,\pi)$.
The two phases  $\xi_Y$   and $\theta_Y\in \mathcal{U}_Y$ can be eliminated 
from the SVD in \eqn{eq:Y} by redefining  in the diagonal matrix of phases $\Phi_Y$ $\varphi_1 \to \tilde \varphi_1 = \varphi_1-\xi_Y-\theta_Y $ 
and $\varphi_2 \to  \tilde \varphi_2 =  \varphi_2-\xi_Y+\theta_Y $, eliminating the $(2_I)$ redundancy.
 For 
$Z$ (and similarly for $K$) 
 the phases $\varphi_2, -(\varphi_1+\varphi_2) \in \Phi_Z$  as well as the 
matrix $V^\dagger_x$, can be dropped since they act on the vanishing entries of $\hat Z$,  
and we can redefine $\xi_Z \to  \tilde \xi_Z =\xi_Z - \varphi_1 $
thus eliminating the redundant  $(1_R, 3_I)$ parameters. Finally the redundant phase in $X$ can be absorbed by 
redefining  $\theta_X \to \tilde \theta_X = \theta_X +\xi_X$. All in all we have
\begin{equation}
\label{eq:Paramt}
Y = \mV_Y^\dagger\, \hat Y\, U_Y,  \quad 
 Z = (V_{Zz}\, V_{Zy})^\dagger  \, \hat Z \, e^{i\xi_Z},  \quad
  K = (V_{Kz} V_{Ky})^\dagger \,  \hat K\, e^{i\xi_K},  \quad 
 X = \hat X\, \Theta\, U_X = \hat X \tilde U_X \Theta,  
\end{equation}
where in the last relation $\tilde U_X =  \Theta U_X \Theta^\dagger  =U_X(\alpha_u, \beta_u-2\theta)$.

Let us now proceed with $SU(3)\times SU(2)\times U(1)$ gauge fixing. 
We can use a full $SU(3)$ transformation  $\mV$ to remove $\mV_Y$ and the $U$ piece of 
the $SU(2)$  transformation $\mU = \Theta U$   to remove $U_Y$, so that  
$Y = \hat Y$. Next we can  use the $\Theta$ part of the   $\mathcal{U}$ transformation  to set 
$X= \hat X \tilde U_X$, and finally we can use the $U(1)$ gauge symmetry to eliminate $\xi_Z$. 
Resuming, after gauging away  $(4_R +8_I)$ components  we have:
\begin{equation}
\label{eq:Pgauge}
Y = \hat Y ,  \qquad 
 Z =(V_{Zz}\, V_{Zy})^\dagger   \, \hat Z ,   \qquad
  K = (V_{Kz} V_{Ky})^\dagger  \,  \hat K\, e^{i\xi_K} \quad (-\pi \leq \xi_K< \pi),  \qquad 
 X = \hat X  \tilde U_X \,. 
 \end{equation}
Note that the $U(1)$ phase $\xi_K$ ranges in the full $2\pi$ interval, and this is because 
 $\gamma_{Kz} \in [-\frac{\pi}{2},\frac{\pi}{2}]$, so that  the effect of $\xi_{K} \to  \xi_{K} +\pi$ 
cannot be reproduced by a $\pi$-shift of  $\gamma_{Kz}$. 
 
In summary, we started with a total of $(14_R, 14_I)$ scalar d.o.f.. By means of the      
$SU(3)\times SU(2)\times U(1)$ symmetry we have gauged away  $(3_R+5_I)+ (1_R+2_I)+ 1_I$ d.o.f. 
leaving $(10_R + 6_I)$ d.o.f.: 5 moduli ($y_1,y_2,z_1,k_1,x_1$), 5 angles of which 
3 of longitude ($\alpha_{Zy},\alpha_{Ky},\alpha_u$) and two of latitude 
($\gamma_{Zz},\gamma_{Kz},$) and 6 phases, one of longitude ($\xi_K$) and 5 of latitude 
($\beta_{Zy},\beta_{Zz},\beta_{Ky},\beta_{Kz},\beta_u$). This  matches the counting of 
the parameters in \eqn{eq:Pgauge}.
Note that since at this stage we have  used  the whole gauge freedom, any other scalar multiplet 
transforming under $SU(3)\times SU(2) \times U(1)$ has to be parametrised in the non-redundant SVD form given in \eqn{eq:Paramt}.


 \twocolumngrid



  \input{AxionFlavor_3.bbl}

\end{document}

%% file: AxionFlavor_3.bbl
%